\documentclass{article}
\usepackage[a4paper, total={6in, 8in}]{geometry}
\usepackage{amsmath}
\usepackage{setspace}
\usepackage{hyperref}
\usepackage{amsmath,amssymb}
\usepackage{amsfonts,amsthm}
\usepackage{graphics}
\usepackage{graphicx}
\usepackage{dcolumn}
\usepackage{color}
\usepackage{bm}
\usepackage[dvipsnames]{xcolor}
\usepackage[normalem]{ulem}
\usepackage[bf]{subfigure}
\usepackage{mathtools}

\title{On the effect of memory on the Prisoner's Dilemma game in correlated networks}

\author{Nastaran Lotfi and Francisco A. Rodrigues}

\begin{document}
\maketitle

\begin{abstract}

Game theory is fundamental to understanding cooperation between agents. Mainly, the Prisoner's Dilemma is a well-known model that has been extensively studied in complex networks. However, although the emergence of cooperation has been investigated before, the influence of memory in its evolution is not well understood. This paper presents a detailed study of cooperation dynamics in which agents have memory. We simulate the evolutionary Prisoner's dilemma game on random, scale-free and networks presenting degree-degree correlation. Through extensive simulations, we show that assortativity can improve cooperation when the temptation to defect increases. Moreover, we show that the inclusion of memory decreases the network structure influence. Our results contribute to understanding the role of the network structure and the player's memory of cooperation. 
\end{abstract}


\section{introduction}
In the last decades, evolutionary theory on complex networks has attracted significant attention from scientists in many areas, from economy to physics~\cite{szabo2007evolutionary,roca2009evolutionary}. Prisoner's dilemma (PD) is a game in which two players acting selfishly will ultimately result in a suboptimal choice for both players. Two players, separated and unable to communicate, must each choose between cooperating with the other or not. Thus, in the game, we have two types of players, i.e., a cooperator (denoted by $C$) and a defector (denoted by $D$)~\cite{robert1984evolution}. Cooperators benefit other individuals at some cost, whereas defectors attempt to exploit such shared resources.  

Initially, the PD was studied on regular graphs~\cite{nowak1992evolutionary}, where subjects played a single game with their neighbours, adopting the strategy of the most successful neighbour for the next round. Next, the PD was adapted to small-world networks, which are obtained through the rewiring of connections in a regular graph~\cite{abramson2001social,masuda2003spatial,santos2005epidemic} or by adding links to a grid structure~\cite{kim2002dynamic,wu2005spatial,ren2007randomness}. Although these analyzes allow a generalization of the traditional PD to graphs, grids, small-world networks, and random graphs do not reproduce most of the properties of real-world networks. Most complex networks have scale-free organization, presenting densely connected nodes called hubs.

Thus, further works adapted the PD to scale-free networks~\cite{santos2005scale,santos2006evolutionary,gomez2007dynamical,poncela2007robustness}. It has been verified that in these networks, cooperation is enhanced in comparison to regular graphs and small-world networks. Indeed, some works investigated the influence of the initial distributions of defectors on the evolution of the game~\cite{chen2008influence}, showing that the initial configurations for defectors can greatly influence the cooperation level and the evolution speed of cooperation. Moreover, the cooperation can be enhanced with the increasing clustering when the initial cooperators are the most connected nodes~\cite{chuang2009prisoner}. In networks with community structure, Chen et al.~\cite{chen2007prisoner} verified that reducing the connections inside the community can promote cooperation as the total links were considered unchanged. The influence of the initial fractions of cooperators on random and scale-free was analyzed in~\cite{poncela2007robustness}. 

All these previous works showed that the structure of the network plays a fundamental role in the evolution of cooperation~\cite{szabo2007evolutionary, perc2017statistical}. However, the player's strategy also influences cooperation. Mainly, players analyze the game and remember their last strategies and the actions of other players. Therefore, players have memory, influencing their planning to maximize their earnings. Although this is an essential ingredient to model cooperation, only a recent paper addressed this issue~\cite{shu2019memory}. In this case, a player applies the memory rule to define her/his action based on her/his historical payoffs. The neighbours' historical optimal procedures are also taken into account to define the player's strategy~\cite{shu2019memory}. The authors verified that historical information promotes cooperation in three classical evolutionary social dilemmas, including the prisoner's dilemma and the snowdrift game. However, only a regular lattice was considered in that study. Therefore, cooperation with memory in heterogeneous networks has not been studied yet.

In this paper, we introduce two new models for capturing the influence of memory in cooperation depending on the players' payoff in complex networks. In the first one, the strategy of player $i$ is defined according to its neighbour's historical payoffs. In the second case, besides the information from the neighbours, he/she considers his/her performance in previous games too. We consider the $m$ previous games in both cases, which define the player's memory.
We assume that players are organized in a complex network whose structure changes from random to scale-free. Degree-degree correlation is also included in the network's structure, enabling us to investigate how the network topology influences the evolution of cooperation. In addition, we modify the neighbour selection process in the traditional PD game. In the standard PD games, neighbours are assumed to be selected randomly, while we add the degree probability for neighbour selection. All models are analyzed considering both methods of neighbour selections to check the influence of neighbours in the evolution of the cooperation. Through extensive simulations, we find that degree-degree correlation and player's memory length are two fundamental ingredients to improve the cooperation between players. However, when we assume that nodes have memory, the influence of network structure is decreased.

The present paper is organized as follows. In section~\ref{sec:method}, we describe the prisoner's dilemma, network construction, and also the memory models. Section~\ref{sec:result} is devoted to presenting and discussing our results. As follows, we conclude our work.

\section{Concepts and methods}\label{sec:method}

\subsection{Traditional prisoner's dilemma}\label{sec:PD}

The simulation of cooperation follows the prisoner's dilemma. Initially, each player is defined as a cooperator ($C$) or a defector ($D$) with the same probability and independently. In the traditional prisoner's dilemma, in each time step, players can only interact with their nearest neighbours with the following rules: if the player is a cooperator, then it receives the payoff $R$ ($S$) as the neighbour is $C$ ($D$). If the player is a defector, the payoff value would be $T$ ($P$) as the neighbour is $C$ ($D$). Due to different values of payoffs, when $T>R>P>S$, the game is named PD. In this paper, values of PD payoffs have been set to $R=1$, $T=b>1$, $P=0$, i.e., no benefit under defectors interactions, and $P-S=\epsilon$~\cite{nowak1992evolutionary,santos2005scale}. Small positive values of $\epsilon\ll1$ have no qualitative differences in the results, where the limit $\epsilon \rightarrow 0^+$ was used in~\cite{nowak1992evolutionary,santos2005scale,gomez2006scale}. Here, we consider the same approach.

At each time step, every node $i$ plays with its nearest neighbours and accumulates the payoffs $p_i$ according to its current state ($C$ or $D$). Next, each player compares its payoff with the payoff of a neighbour chosen at random. In our study, the random selection of the neighbours is made in two ways: (i) at random or (ii) according to the node degree. In this way, the neighbours with a higher degree get a higher chance of being selected. Player $i$ will keep its strategy if $p_i>p_j$ or will copy $j$'s strategy with probability 
$$
P_{i\rightarrow j}=\frac{1}{max(k_i,k_j)*b}(p_j-p_i)\quad \text{ if } p_i<p_j.
$$
where $k$ is the node degree (number of connections).
The state of the nodes is defined by the vector $s_i (i=1,..., N)$, whose entries are equal to 1 if node $i$ is a cooperator or 0 if it is a defector. As a result, an instantaneous fraction of cooperators at time $t$ can be defined as $C(t)=N^{-1}\sum_i s_i(t)$. After the transient time $t_0$ (which is large enough to make the mean value of $C$ stationary), we get the average number of cooperators ($\langle C\rangle$) for each parameter value $b$.

\subsection{The prisoner's dilemma with memory}

The simulation of the PD with memory considers two classes of nodes, cooperator ($C$) and defector ($D$). At the beginning of the process, each node is classified in one of these two states. The initial fraction of cooperators is a parameter of the simulation. Each node selects a neighbour and changes its state to maximize its gains.

To verify how memory affects cooperation, we consider two different possibilities. In the first one, at each game, a player can keep its last strategy (C or D) or change it according to the neighbour's actions. In this case, a player analyzes the payoff obtained by its neighbours in previous games. On the other hand, a player can consider its previous strategies to decide to cooperate or defect along with the neighbors history. To simulate these two different scenarios, we consider these two models:
\begin{enumerate}
    \item In model $A$, each node sets its last state according to its neighbour's memory. To do so, node $i$ will decide to changed its strategy if $p_i<\overline {p}_{j}^{m}$ with probability:
\begin{equation}\label{eq:1}
    P_{i\rightarrow j}=\frac{1}{max(k_i,k_j)*b}(\overline {p}_{j}^{m}-p_i),
\end{equation}
where $\overline {p}_{j}^{m}$ is the neighbours' average payoff over the previous $m$ games. The neighbour $j$ is selected with a probability that depends on its degree $k_j$ (preferential selection according to the number of connections) or uniformly at random.
\item In model B, we adopt the same configuration as in model A. However, we add the memory influence of member $i$ too. In other words, we assume that the node has some background from its previous games and also its neighbors history. As a result, node $i$ will decide to changed its strategy if $\overline {p}_{i}^{m}<\overline {p}_{j}^{m}$ with the probability:
    \begin{equation}\label{eq:2}
     P_{i\rightarrow j}=\frac{1.0}{max(k_i,k_j)*b}(\overline {p}_{j}^{m}-\overline {p}_{i}^{m}).
    \end{equation}
\end{enumerate}
In both models, the traditional PD rules are applied until reaching the memory step $m$, and after that, the new defined models are operated. Studying neighbour's power in the decision making of the members, for each of the models mentioned above, we examine two types of neighbour selection. At first, we consider a regular random selection of neighbours in which there is no priority for selection. In this way, neighbours have the same chance to be selected. In the second method, we examine the influence of selecting neighbours according to their degrees. In this case, higher degree nodes are more likely to be selected.

\subsection{Network construction}

Many works have verified that most real-world networks are heterogeneous, such that the probability distribution of the number of connections (degree) follows a power law. These networks are called scale-free networks~\cite{newman2003structure,boccaletti2006complex}. Exponential graphs cannot describe these networks, that present a Poisson-like degree distribution (e.g. random networks)~\cite{barabasi1999emergence,barabasi1999mean}. There are many models to generate scale-free and random networks. For example, Gomez et al.~\cite{gomez2006scale} proposed a new model to get a smooth transition from random to scale-free networks. To construct the network, we start with fully connected $m_0$ nodes ($m_0=3$ in our case), and the rest of the nodes ($N-m_0$) are disconnected. At each time step, a new node is added to the network. With probability $\alpha$, this node is connected to any of the $N-1$ nodes in the network. With probability $(1-\alpha)$, the node establishes a link according to the preferential attachment strategy, i.e., a node's probability of receiving a new link is proportional to its degree. For each node, this process is repeated $\overline {k}$ times, where $\overline {k}$ is the mean degree of the network, and all nodes from a set of ($N-m_0$) are considered. Notice that the transition from random to scale-free networks is controlled by the parameter $\alpha$, where $\alpha=0$ for pure random networks and $\alpha=1$ for a scale-free organization. 

We also consider assortative networks, which are given by degree-degree correlations. Assortativity is defined as the tendency of nodes with a similar degree to be connected~\cite{newman2002assortative}. In other words, the assortativity coefficient is measured as the Pearson correlation coefficient $(r)$ of the node degree at the end of each edge. Positive values of $r$ indicate that nodes of similar degrees are connected. On the other hand, negative values suggest that hubs tend to connect to low-degree nodes. The assortativity is calculated by\cite{newman2018networks}:
\begin{equation}\label{Eq:assortativity}
r = \frac{\sum_{i=1}^N \sum_{j=1}^N \left( A_{ij} - \frac{k_i k_j}{2M}\right) k_i k_j}{\sum_{i=1}^N \sum_{j=1}^N \left( ki\delta(i,j) - \frac{k_i k_j}{2M}\right) k_i k_j}.
\end{equation}
where $A_{ij}$ is equal to one if there is a connection between nodes $i$ and $j$ or equal to zero, otherwise.

To change from random to assortative (disassortative) scale-free networks, we modify the algorithm~\cite{noh2007percolation}. In this case, two random links are selected and rewired if the network's total assortativity (dissassortativity) increases (decreases). Otherwise, the system keeps its first configuration. The node degree is preserved during this process.

With the rules mentioned above, constructed networks will be connected as each node will have at least one connection. For the rest of the paper, the size of the undirected networks and the mean degree of the nodes is considered as $N=2000$ and $\overline {k}=4$, respectively. Three values for $\alpha$ are considered here: (i) $\alpha=1$ for random networks, (ii) $\alpha=0.5$ for networks between random and scale-free structures, and (iii) $\alpha=0.3$ for scale-free networks. We verify that as $\alpha$ is close to zero, the obtained results are noisier, although the behaviour trend is the same.

\section{Results}\label{sec:result}

\begin{figure*}
    \centering
    \includegraphics[width=\textwidth,height=7.5cm]{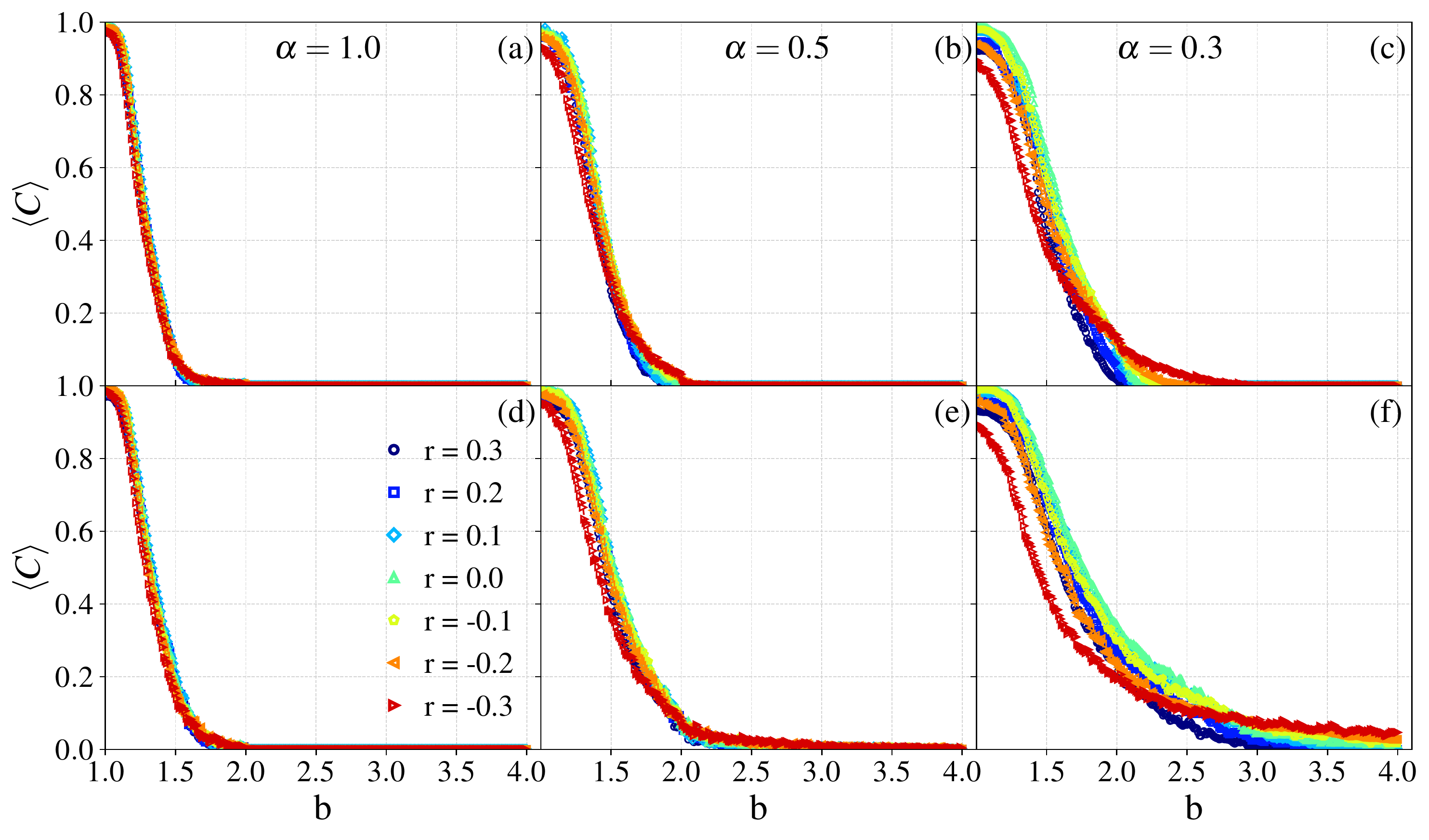}
    \caption{Fraction of cooperations according to the payoff $T=b$ (the temptation to defect) for random ($\alpha=1$), half-random ($\alpha=0.5$) and scale-free ($\alpha = 0.3$) networks. Results for the uniform neighbour selection are shown in (a)-(c), whereas the neighbour sampling in terms of the node degree is depicted in (d)-(f).}
    \label{fig:normal_alpha}
\end{figure*}

Initially, we consider the traditional prisoner's dilemma model without memory. Each player defines its strategy by randomly selecting a neighbour to compare its payoff. This neighbour selection can be made at random or according to the neighbour's degree. In figures~\ref{fig:normal_alpha} (a) and (d) ($\alpha=1$), we show the results for random networks. As we can see, the level of assortativity does not influence the fraction of cooperators. Moreover, the way we select the neighbours to interact with does not affect the level of cooperation. This lack of cooperation influence is an expected result because random networks are homogeneous, and hubs are absent.

When we increase $\alpha$ to 0.5 (figures~\ref{fig:normal_alpha} (b) and (e)), we can see that assortativity starts to play an essential role in cooperation dynamics. Indeed for higher values of $\alpha$, when the networks turn to become scale-free networks (figures~\ref{fig:normal_alpha} (c) and (f)), for small values of $b$, the level of cooperation is higher for non-assortative and assortative networks. On the other hand, when $b$ is increased, the cooperation is enhanced in disassortative networks. Moreover, selecting the neighbours according to the degree improves the cooperation in disassortative networks. For assortative networks, the cooperation decreases faster as the value of the parameter $b$ is increased. Therefore, assortativity plays an essential role in cooperation, but its influence is not trivial. It depends on the defection temptation, which can improve cooperation if it is more prominent in disassortative networks. 

\begin{figure*}
    \centering
    \includegraphics[width=\textwidth,height=7.5cm]{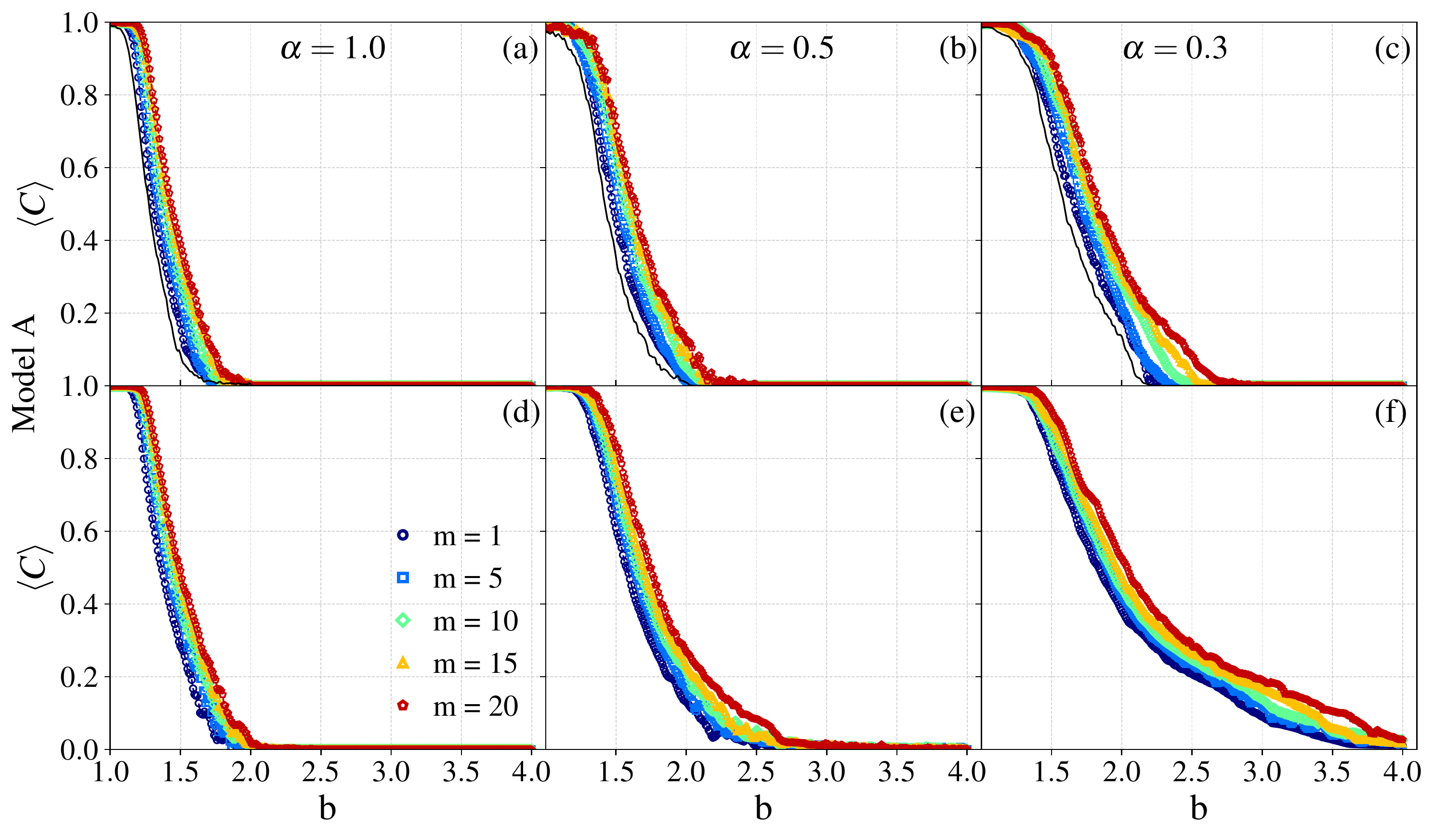}
    \caption{Memory influence in PD in complex networks by assuming the model $A$ with $r=0$. Results for the uniform neighbour selection are shown in (a)-(c), whereas the neighbour sampling in terms of the node degree is depicted in (d)-(f). }
    \label{fig:memory-normal-A}
\end{figure*}

We repeated the same analysis by including the player's memory to verify how the information about previous games influences the evolution of cooperation. In figure~\ref{fig:memory-normal-A}, we show the results for the traditional prisoner's dilemma game with memory. We consider model A, in which each player follows the information about the previous games of their neighbours to define their current strategy. The plots show that increasing the memory length implies the growth of cooperation, independently of the value of $\alpha$, i.e., for random and scale-free networks. Also, the neighbour selection according to the degree tends to increase the level of cooperation, as shown in figure~\ref{fig:memory-normal-A}(d-f). 

\begin{figure*}
    \centering
    \includegraphics[width=\textwidth,height=7.5cm]{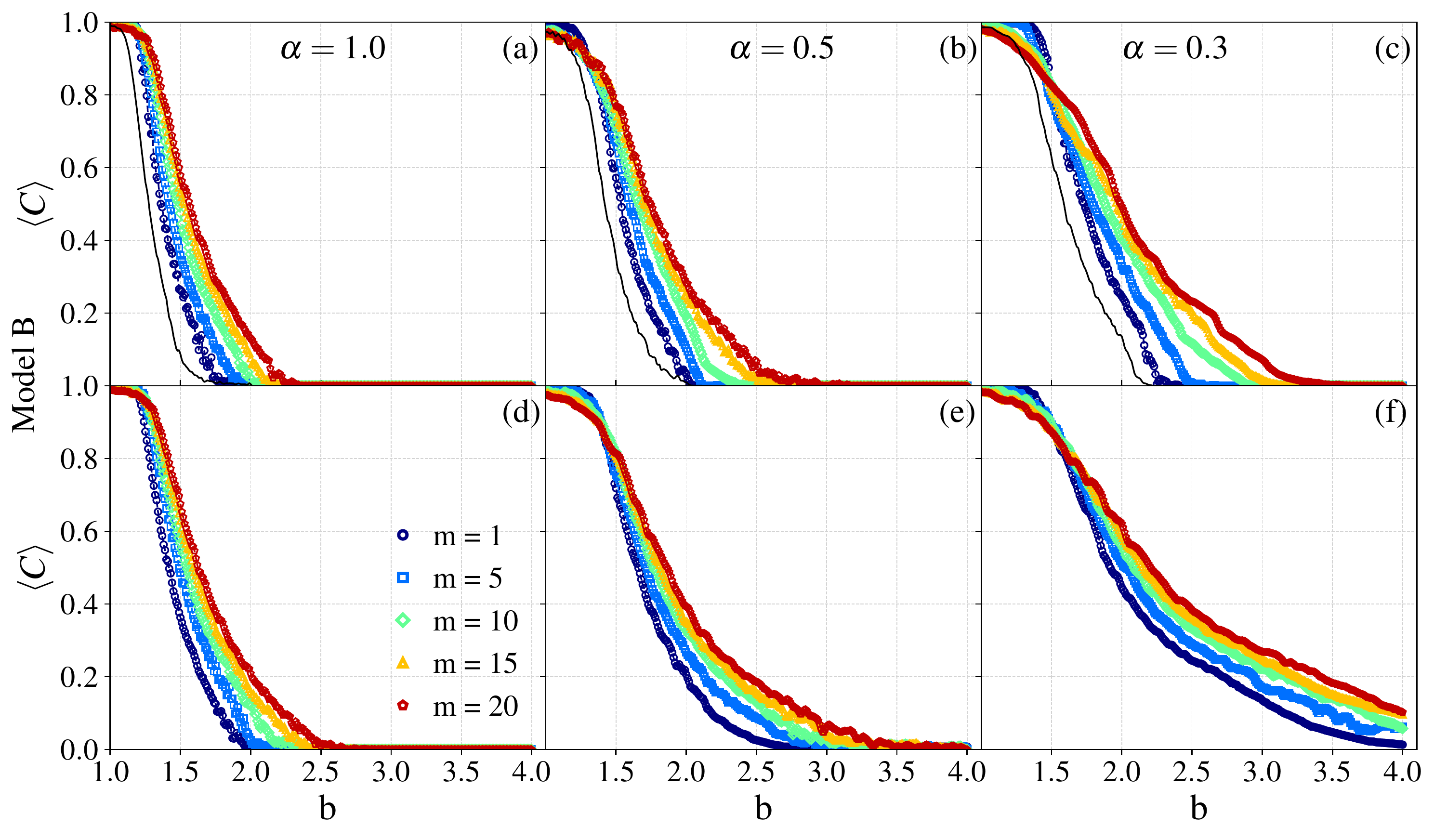}
    \caption{Memory influence in PD in complex networks by assuming the model $B$ with $r=0$. Results for the uniform neighbour selection are shown in (a)-(c), whereas the neighbour sampling in terms of the node degree is depicted in (d)-(f).}
    \label{fig:memory-normal-B}
\end{figure*}

For the case of model B, in which each player adds its historical game recordings to define its strategy, the results are shown in figure~\ref{fig:memory-normal-B}. The memory helps improve the cooperation only for larger values of $T=b$. Indeed, for scale-free networks ($\alpha=0.3$), we can see that the cooperation is enhanced only for $b>1.5$. Again, the neighbour selection by considering the node degree raises the level of cooperation, as shown in figure~\ref{fig:memory-normal-B}(d-f). 

\begin{figure*}
    \centering
    \includegraphics[width=\textwidth,height=11cm]{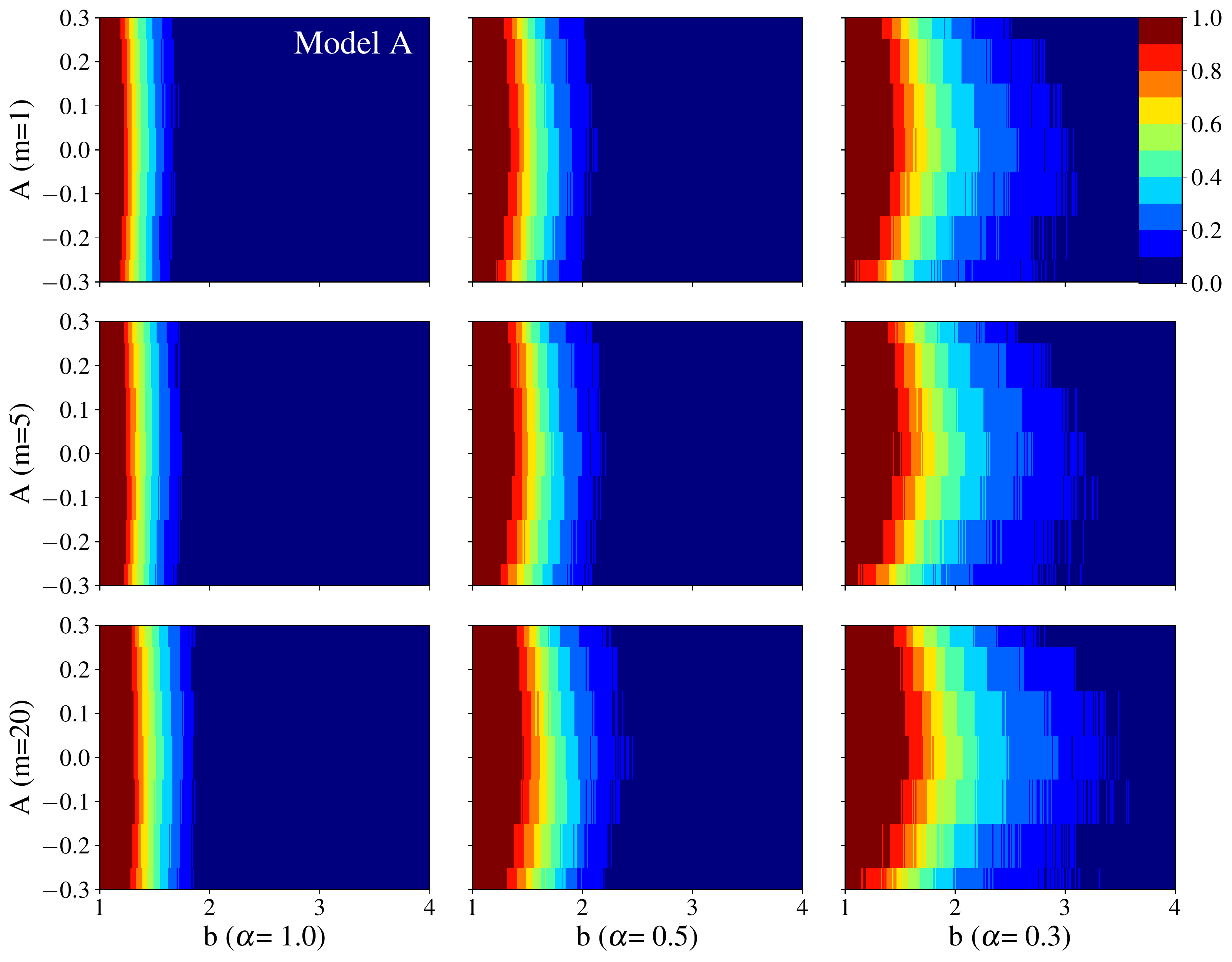}
    \caption{Level of cooperation in degree-degree correlated networks for model A. We consider three memory lengths (m=1, 5, and 20) and networks topologies (random ($\alpha=1$), half-random ($\alpha=0.5$) and scale-free ($\alpha=0.3$)).Colors represents the fraction of cooperators $\langle C\rangle$.}
    \label{fig:memory-modelA}
\end{figure*}

\begin{figure*}
    \centering
    \includegraphics[width=\textwidth,height=11cm]{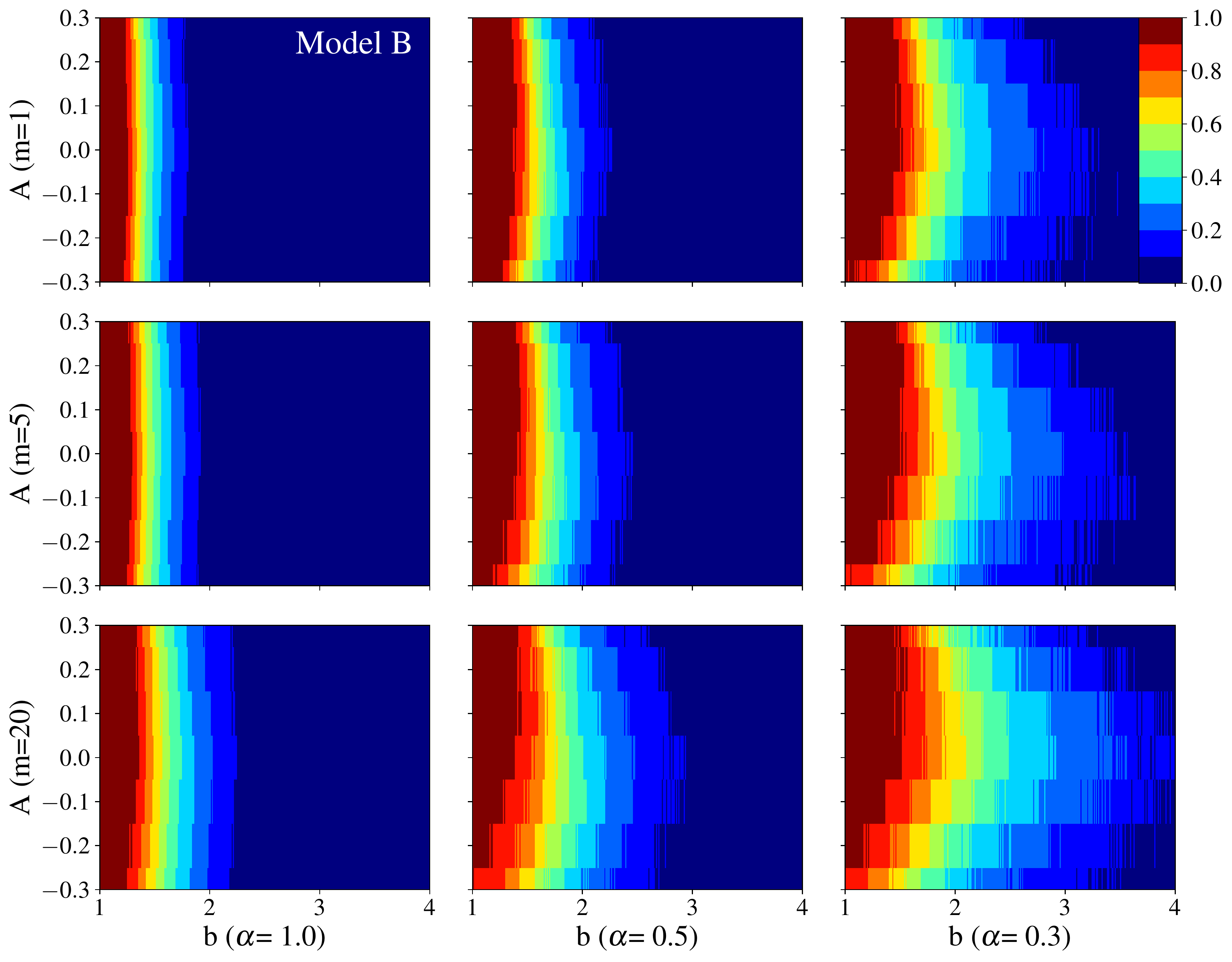}
    \caption{Level of cooperation in degree-degree correlated networks for model B. We consider three memory lengths (m=1, 5, and 20) and networks topologies (random ($\alpha=1$), half-random ($\alpha=0.5$) and scale-free ($\alpha=0.3$)).Colors represents the fraction of cooperators $\langle C\rangle$.}
    \label{fig:memory-modelB}
\end{figure*}

To verify the effect of memory and assortativity altogether, we simulate the prisoner's dilemma in degree-degree correlated networks. The results for model A, which considers the information about the neighbour's payoff, are shown in figure~\ref{fig:memory-modelA}. We consider three memory lengths ($m=1$, $5$ and $20$) with different assortativity values for three types of networks ($\alpha = 1.0, 0.5$ and $0.3$). Neighbours are selected according to their degree. In all cases, increasing the memory length increases the level of cooperation. However, non-correlated scale-free networks present the highest fraction of cooperators (see figure~\ref{fig:memory-modelA}(c)). For model B, results are in figure~\ref{fig:memory-modelB}. Again, uncorrelated networks with the highest memory length present the highest level of cooperation. 

Compared with the previous results, we can see that memory length decreases the influence of degree-degree correlation on the level of cooperation. Moreover, the player strategy (models A or B) does not impact the final fraction of cooperators significantly. Even so, scale-free networks are the topology that most improves cooperation.

\section{Conclusion}
This paper has studied the prisoner's dilemma in correlated networks. We have verified that assortativity plays a vital role in defining cooperation. For small values of the temptation payoff, assortativity impairs cooperation in scale-free networks. On the other hand, cooperation is more considerable when this payoff increases than in non-correlated or disassortative networks. The network structure's effect is weak when we include the player's memory, who can consider her/his last games or the last neighbours' games. Cooperation impairs when we consider small memory lengths, whereas cooperation improves for longer memory lengths. Also, scale-free networks promote the highest level of cooperation when the effect of memory is present.

Therefore, we have verified that non-trivial patterns of connections and the player's strategy are two ingredients to be considered to improve the cooperation between agents. These results can be continued considering other patterns of connections, like community structure and the presence of cycles in the network. Moreover, the extension for multilayer networks~\cite{kivela2014multilayer,boccaletti2014structure}, made of connected networks, is also a promising analysis since these networks represent social interactions with more details than traditional single-layer networks.

\section{acknowledgement}
N. Lotfi is thankful to the FAPESP (grant with number 2020/08359-1) for the support given to this research.


\bibliographystyle{ieeetr}
\bibliography{bib}
\end{document}